\documentclass[%
 aip,
 jmp,%
 amsmath,
 amssymb,
preprint,%
]{revtex4-1}

\usepackage{float} 

\usepackage{graphicx}
\usepackage{bm}

\begin{document}


\title{Quantitative analysis of timing in animal vocal sequences }

\author{Florencia Noriega}
\email[]{florencia.noriega@code.berlin}

%
\affiliation{
Chair for Network Dynamics, Center for Advancing Electronics Dresden (cfaed) and 
Institute for Theoretical Physics, TU Dresden, 01062 Dresden, Germany
}
\affiliation{CODE University of Applied Sciences, Lohm\"uhlenstra{\ss}e 65, 12435 Berlin, Germany}
\affiliation{
Network Dynamics, Max Planck Institute for Dynamics and Self-Organization, 37077 G\"ottingen, Germany
}%

\author{Adolfo Christian Montes-Medina}
\affiliation{%
Instituto de Investigaciones en Ecosistemas y Sustentabilidad (IIES), Universidad Nacional Aut\'onoma de M\'exico, Morelia, Michoac\'an, M\'exico.
}%

\author{Marc Timme}%
\affiliation{
Chair for Network Dynamics, Center for Advancing Electronics Dresden (cfaed) and 
Institute for Theoretical Physics, TU Dresden, 01062 Dresden, Germany
}
\affiliation{
Network Dynamics, Max Planck Institute for Dynamics and Self-Organization, 37077 G\"ottingen, Germany
}%
\affiliation{Cluster of Excellence Physics of Life, TU Dresden, 01062 Dresden, Germany}

\date{\today}%

\begin{abstract}

Timing features such as the silence gaps between vocal units -- inter-call intervals (ICIs) -- often correlate with biological information such as context or genetic information. 
Such correlates between the ICIs and biological information have been reported for a diversity of animals. 
Yet, few quantitative approaches for investigating timing exist to date.
Here, we propose a novel approach for quantitatively comparing timing in animal vocalisations in terms of the typical ICIs.
%
As features, we use the distribution of silence gaps
parametrised with a kernel density estimate (KDE) and
compare the distributions with the symmetric Kullback-Leibler
divergence (sKL-divergence).
We use this technique to compare timing in vocalisations of two frog species, a group of zebra finches and calls from parrots of the same species. 
As a main finding, we demonstrate that in our dataset, closely related species have more similar distributions than
species genetically more distant, 
with sKL-divergences across-species larger than within-species distances. 
Compared with more standard methods such as Fourier analysis, 
the proposed method is more robust to different durations present in the data samples, 
flexibly applicable to different species and easy to interpret.
Investigating timing in animal vocalisations may thus contribute to taxonomy, support conservation efforts by helping monitoring animals in the wild and may shed light onto the origins of timing structures in animal vocal communication.

\end{abstract}

\maketitle

\section{\label{sec:intro} Introduction}

Bioacoustic signals contain information useful in diverse disciplines such as conservation, ecology, evolution and ethology. 
%
These signals are often studied through spectrograms 
from which vocal units --whale calls, dolphin whistles, parrot notes, frog pulses-- are identified and classified. 
Vocal units can indicate food availability \cite{mahurin2008chick, slocombe2006food},
individual \cite{freeberg2012geographic, sayigh1999individual, charrier2009vocal} or group identity \cite{ford1989acoustic, attard2010vocal, arcadi1996phrase} and 
warn other group members of the presence of a predator \cite{seyfarth1980vervet, casar2012evidence, freeberg2012geographic}. 
Through the spectral domain, vocal units often convey biological information among animals, yet, this is not the only domain nor always the most relevant one \cite{nevo1985evolutionary, bohn2008syllable}.

\begin{figure*}[!htbp]
\center
\includegraphics[width=\textwidth]{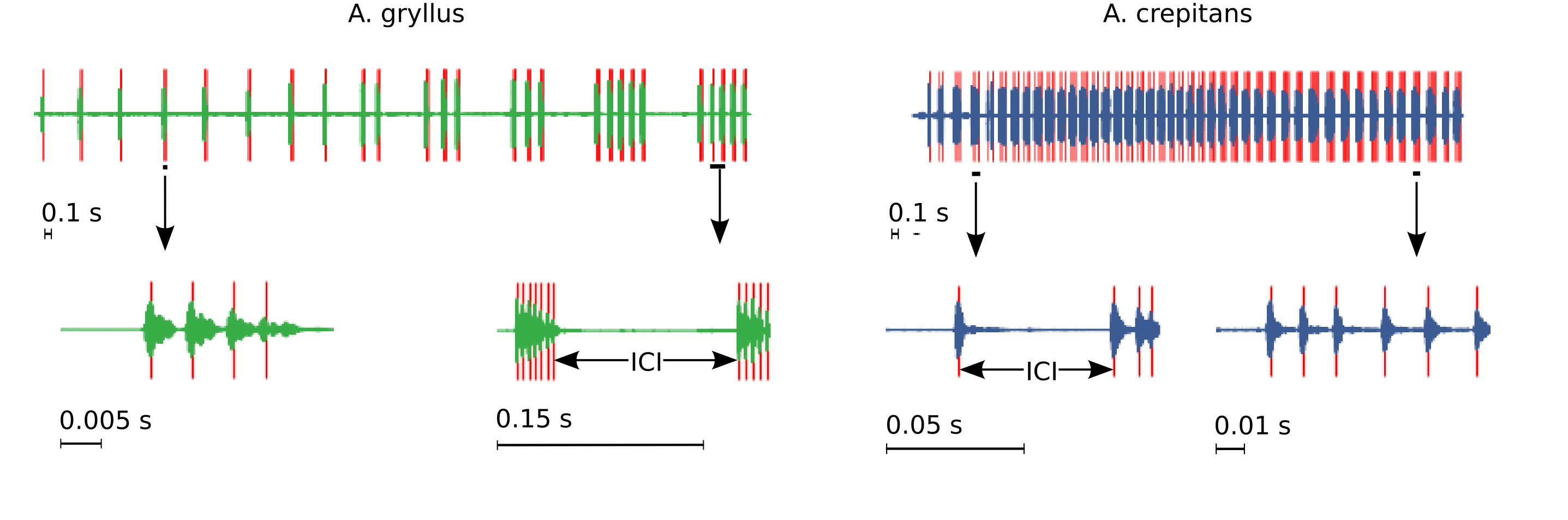}
\caption{
{\bf Mating songs of two species of cricket frogs. }
Waveforms with pulses of frogs (a) {\it Acris gryllus} and (b) {\it Acris crepitans}. 
Both songs are made out of pulses hierarchically grouped in time.  
However, the way the pulses group -- the inter-pulse intervals and the inter-group intervals -- are different for both frogs. 
{\it Acris gryllus} has 
inter-pulse intervals of ca.~\mbox{0.005 s} and 
inter-group intervals of ca.~\mbox{ 0.6 s}
and 
{\it Acris crepitans} has 
inter-pulse intervals of ca. 0.01 s and 
inter-group intervals of ca. 0.1 s. 
Red lines indicate the position of the pulses, located thresholding the signal at 60\% with a minimal inter-peak distance 
of 80 samples for {\it Acris gryllus}
and of 100  samples for {\it Acris crepitans} \cite{micancin2014allometric}. 
Thresholding parameters were chosen ad hoc through visual inspection. 
Recordings from \cite{micancin2014allometric}.
}
\label{fig:frog-call}
\end{figure*}

\begin{figure}[!h]
\centering
\includegraphics[width=0.6\linewidth]{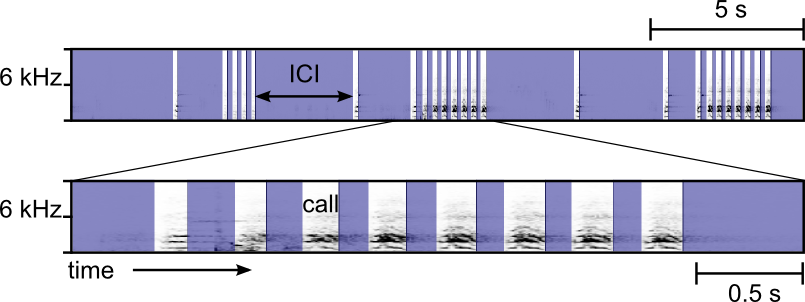} 
\put(0,0){a}

\includegraphics[width=.6\linewidth]{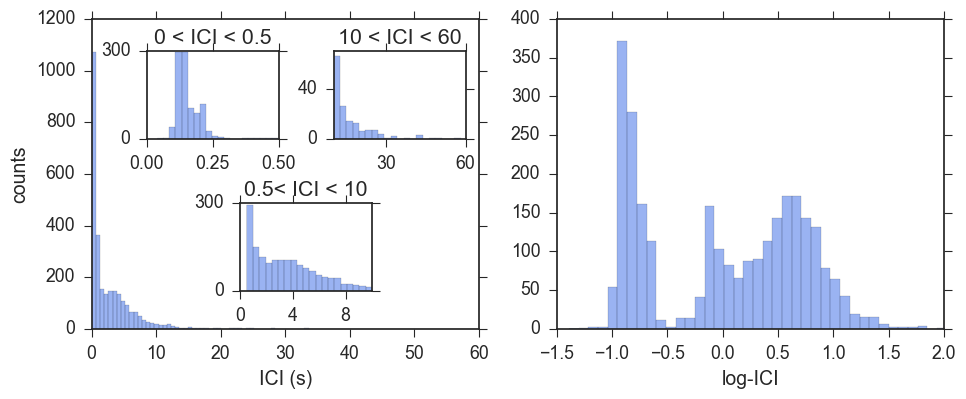}
\put(0,0){b}
\put(-120,0){c}
\caption{{\bf Inter-call intervals (ICIs) range along several orders of magnitude.} 
{\bf a}, Spectrogram with parrot vocalisations (calls) and ICI indicated with blue segments. 
 Distribution of ({\bf b}) ICIs and of the ({\bf c}) logarithm base ten of the ICIs (log-ICIs).
}
\label{fig:spectro}
\end{figure}

Biological information can also be coded in the temporal domain. 
Cricket frogs, for instance, select mates based on the temporal structure of their mating songs and not on the spectral characteristics \cite{nevo1985evolutionary}. 
As a consequence, mating songs from different species of cricket frogs are temporally distinctive (Fig.~\ref{fig:frog-call}). 
In addition to frogs, other species also show correlations between the timing of their vocalisation and biological factors. %
Calling rates, for instance, are affected by diverse contextual factors that include 
environment \cite{simard2010depth}, 
food availability \cite{evans1994food}, 
the presence of other group members \cite{evans1994food, wells1984vocal, matos2005performing}, 
group size \cite{payne2003elephant} and
anthropogenic noise \cite{sun2005anthropogenic, kaiser2009effect, potvin2011geographically}.
Like the calling rates, the duration of silences between consecutive calls, or ICIs  (Fig.~\ref{fig:spectro}), 
also provide information on the levels of vocal activity; a large density of short ICIs is associated with high vocal activity whereas a large density of long ICIs is associated with low vocal activity.
The ICIs can be characteristic of an individual \cite{cranford1999sperm, gero2016individual}, a species or a population \cite{stark1998quantitative, mann2009comparative, weilgart1997group}.
Timing is an important domain of animal vocal communication and investigating it may shade light into how temporal structures evolved and how animals use them to communicate.
 
%
Despite of this evidence for the importance of temporal domain for animal communication, there are few quantitative approaches for investigating timing and particularly the ICIs. 
Because the ICIs can range over several time scales (Fig.~\ref{fig:frog-call} and  \ref{fig:spectro}), 
central statistic measures like the mean or the median are of little help. 
%
One way to handle the different time scales is to define hierarchical elements such as pulses, pulse groups, clicks, click groups and bouts \cite{nevo1985evolutionary, bohn2008syllable}. 
However, these semantic categories are often not distinguishable in reality. 
For example, in Fig.~\ref{fig:frog-call}a, the song of frog {\it A. gryllus}  
has no one inter-group interval size, 
but they range between 0.15 s and 0.6 s. 
Instead of defining pulses and pulse groups, a less biased approach would be to handle all the same regardless of whether they are part of an ad hoc pulse cluster or not and see what kind of structures emerge. 
%
%

Here we propose an approach for quantitatively comparing timing patterns in animal vocalisations in terms of the typical silence gaps between consecutive vocal units.  
As features, we use the distribution of silence gaps parametrised with a kernel density estimate (KDE) and 
compare the distributions with the symmetric Kullback-Leibler divergence (\mbox{sKL-divergence}). 
We use the method to compare three datasets with vocalisations of different animals: 
two frog species, 
a group of four zebra finches and 
vocalisations from different males of a parrot species. 
Given that the animals within the datasets are genetically closer than the animals across the datasets we would expect the across dataset distances to be larger than the within dataset distances. 
Finally, we contrast the proposed method with Fourier analysis. 
Quantifying differences in the typical ICIs of animals can contribute to understanding animals more thoroughly and support conservation efforts.

\section{Methods}

The proposed method is aimed for investigating timing of vocal units in general and regardless of whether the vocal units are dolphin whistles, parrot notes, frog pulses or any other vocal unit (see \cite{kershenbaum2014acoustic} for a review on different ways of defining vocal units). 
So, for simplicity, we will use 
the term call to refer to the vocal units (Fig.~\ref{fig:spectro}) and 
we the term inter-call interval (ICI) to refer to the silence gaps.

The method consist of two parts: feature extraction and comparison. 
In the feature extraction part, distributions of ICIs are parametrised with a Gaussian KDE and in the comparison part, distances between the distributions are  quantified with the \mbox{sKL-divergence}. 
The details of the datasets are summarised in Table \ref{tab:data} 
 and the methods are presented below.


\begin{table}[H]

 \resizebox{\textwidth}{!}{

\begin{tabular}{ p{1.2cm} p{0.091cm} p{6.5cm} p{0.091cm} p{9.1cm} }
\hline
{\bf Dataset} & & {\bf Biological description } &	& {\bf Recordings and annotations}\\
\hline
frogs
&&
Two species of cricket frogs {\it Acris crepitans} and {\it Acris gryllus}. 
&&
Two recordings, one per species, of 11 s and 12 s long form \cite{micancin2014allometric}. 
Only one animal vocalising in each recording.
Recordings annotated with frog pulses by peaking the signal (Fig.~\ref{fig:frog-call}).
\\
\\
zebra finches (zf)
&& 
Four female zebra finches {\it Taeniopygia guttata} recorded on two consecutive days in the lab in the same acoustic and visual space.  
&&
Two annotated recordings from \cite{stowell_2016} labelled as zf2 and zf3.
Annotations indicate onset times, duration of the calls and signaller ID. 
In the present study we were not interested in the individual ID, so we disregard this information.  
The ICIs of this dataset may come from calls produced by the same or two different individuals. 
Negative ICIs occurring when calls overlapped where filtered out of the dataset to avoid singularities in the logarithm. 
\\
\\
parrots
&&
Nesting vocalisations from wild males Lila-crowned parrots ({\it Amazona finschi}) recorded in the Chamela-Cuixmala Biosphere Reserve, Mexico. 
During the beginning of nesting period male parrots display a stereotypical behaviour in which they approach to nest cavity, where the female is inside incubating eggs, and call to her until she comes out of the nest cavity and feed her\cite{renton1999nesting}. 
Recordings obtained opportunistically when the male was found vocalising and stopped when female came out of the nest cavity. 
&&
Only one male parrot vocalising per recording. 
Females varied in time to response to her males, so male calls varied in duration and number of calls. 
Recordings with less than 30 calls and shorter than three minutes were disregarded. 
Only males with two or more recordings were considered. 
This yielded a dataset with 12 recordings from four individuals, two to four recordings per parrot, recording duration ranged between three and 12 minutes, with 46 to 186 calls per tape. 
Vocal units annotated manually from spectrograms using Raven version 1.4 and Audacity \cite{montes2016contextual}. 
Recordings named with nest parrot ID – K3, GA, FH, FC – followed by a recording ID – given by a capital R and an index – separated by a dash. 
So, label K3-R1 corresponds to recording one of male K3.
\\
\hline
\end{tabular} }
\caption{{\bf Datasets.} 
Dataset name (first column), 
description of the context in which the recordings were taken (second column)
and description of the annotation of the recordings (third column). }

\label{tab:data}

\end{table}

\subsection{Feature extraction: distribution of ICIs}
\label{sec:quant}

\begin{figure}[H]
\centering
\includegraphics[width=.49\linewidth]{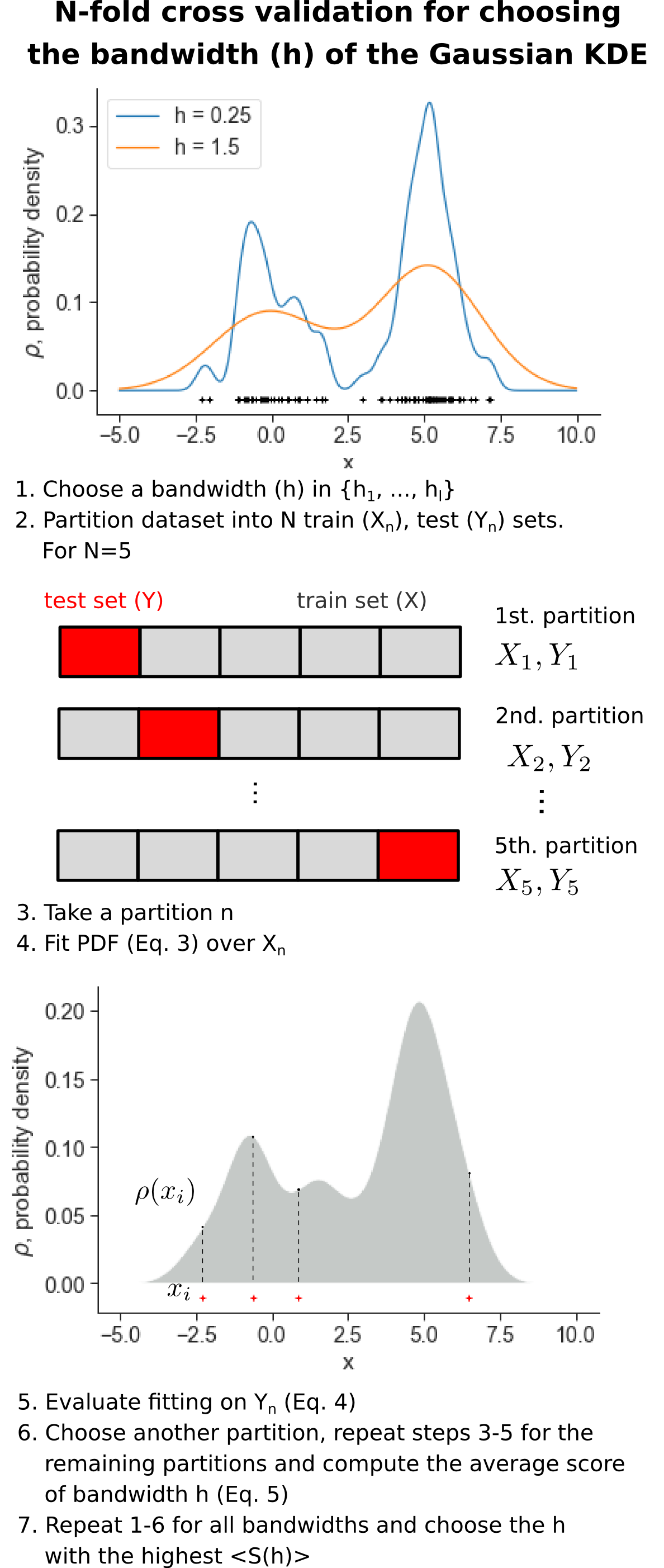}
\caption{
{\bf Choosing bandwidth ($h$) of the KDE with cross-validation. }
Use cross-validation to choose the estimate $\hat{\rho} _h$ with the highest cross validation score. 
}
\label{fig:CV}
\end{figure}

Distributions of ICIs are typically broad; ranging from values of a fraction of a second to ICIs of several seconds. 
Figure ~\ref{fig:spectro}, for instance, shows 
ICIs as small as \mbox{ca. 0.04 s} as well as ICIs more than a thousand times larger than \mbox{0.04 s.}  
To comprise the structure of the distribution of ICIs along this wide range covering several time scales, we take the logarithm of the ICIs (log-ICI).
The logarithm scales the ICIs according to their magnitude, increasing the resolution at small scales and decreasing the resolution at large scales, thereby enabling to visualise the typical time scales of the ICIs.
Given the distribution of log-ICIs $\rho$, 
the probability of having an ICI $t$ larger than $a$ and smaller than $b$, $p_{t \in [a, b]}$ is given by
%
%
\begin{equation}
p_{t \in [a, b]} = \int \limits _{a} ^{b} \frac{\rho (\log (t))}{t \cdot \ln(10)} dt. 
\label{eq:log-prob}
\end{equation}
The expression above is obtained by applying the chain rule to the integral of the probability density function of the log-ICIs, $x = \log _{10}(t)$. 
%
The kernel density estimate (KDE) is a method for fitting the probability distribution $\rho$ of a continuous variable $x$. 
It is pretty much like a histogram but without the binning, thereby yielding a smoother distribution. 
The method consist of summing kernel functions $K_h$ centred the $m$ points of the sample ${x_i}$ we wish to parametrise. 
So, the KDE $\hat{\rho}_{h}$ of $\rho$ is
\begin{equation}
\hat{\rho}_{h} (x) = 
\frac{1}{ m h } \sum \limits _{j=1} ^{m} K_h \big(x-x_i \big),
\label{eq:kde-estimate}
\end{equation}
where $h$ is the bandwidth of the Kernel. 
The estimate is highly dependent on $h$ and below we detail how this parameter is chosen, but before let us talk about the kernel. 
Commonly a Gaussian kernel is used, in which case the bandwidth corresponds to the standard deviation of the normal distribution  \cite{gramacki2018nonparametric}
\begin{equation}
\hat{\rho}_{h} (x) = 
\frac{1}{ \sqrt{2 \pi } m h } \sum \limits _{j=1} ^{m} \exp \Bigg(\frac{(x-x_i)^2}{2h^2} \Bigg). 
\label{eq:gkde-estimate}
\end{equation}

To choose the bandwidth we carry out a cross validation scheme over the estimate $\hat{\rho}$ (Fig.~\ref{fig:CV}). 
Cross validation is a technique for evaluating data fitted models, by partitioning data into training and testing subsets. 
In the case of a KDE, we use a train subset $X$ to fit $\hat{\rho} _{h}$ 
and a test subset $Y$ for evaluating $\hat{\rho} _{h}$ as the likelihood of the samples  of $Y$ under the KDE $\hat{\rho} _{h}$, $\hat{\rho} _{h} (x)$, for $x$ in $Y$. 
Because the probabilities $\hat{\rho} _{h} (x)$ are typically very small values it is common to use the log-probabilities instead. 
This increases the resolution of likelihood without affecting its maximum, given the logarithm is a monotonically increasing function. 
Therefore, we evaluate the estimate $\hat{\rho} _h$ under the samples of $Y$, through the score $S(h)$ given by 
\begin{eqnarray}
S(h) = \sum _{x_i \in Y} \log (\hat{\rho}_h (x_i)).
\label{eq:kde-score}
\end{eqnarray}
Naturally, the score $S(h)$ depends on the partition of the dataset. 
To counteract the effects of the partition we perform a N-fold cross validation (Fig. \ref{fig:CV}). 
This consist of partitioning the sample $N$ times, 
fitting $\hat{\rho} _h$ and evaluating it for each of the $N$ partitions, and 
finally choosing the $h$ with the highest average score, 
\begin{eqnarray}
\langle S(h) \rangle = \frac{1}{N} \sum \limits _{n=1} ^{N} S_n(h),
\label{eq:kde-av-score}
\end{eqnarray}
where $S_n$ is the score for the $n$-th partition. 
In a N-fold cross validations
there are $N$ training subsets $X_n$,
with $\frac{N-1}{N}$ of the samples, 
with their corresponding $N$ testing subsets $Y_n$, 
with $\frac{1}{N}$ of the samples. 
The same way as was done before, 
subset $X_n$ is used for fitting the estimate $\hat{\rho}_h$ and 
subset $Y_n$ is used for evaluating the estimate $\hat{\rho}_h$. 
We use the KDE implemented scikit-learn's \cite{pedregosa2011scikit} Kernel Density Estimate with a N-fold cross validation with $N=10$, trying ten bandwidths logarithmically (base 10) increasing from 0.1 to one.  
Besides the Gaussian, other kernels can be used with \mbox{Eq. \ref{eq:kde-estimate}} such as cosine, Epanechnikov, linear, exponential and tophat. 
Like was done for choosing the bandwidth, we evaluate the kernels with  ten fold cross validation 
and obtain that Gaussian yield the best estimates and so we use this kernel. 

\subsection{Comparing distributions: sKL-divergence}

\begin{figure}[h]
\centering
\includegraphics[width=0.65\linewidth]{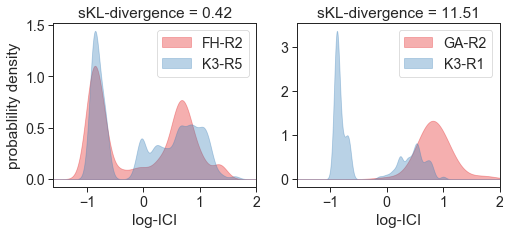}
\put(-120,0){\bf a}
\put(0,0){\bf b}
\caption{
{\bf Comparing distributions of inter-call intervals (ICIs).} 
Probability distribution of the log-inter-call intervals (log-ICIs) of 
(a) tape R2 of parrot FH and tape R5 of parrot K3 
and (b) tape R2 of parrot GA and tape R1 of parrot K3. 
Probability distribution fitted with a Gaussian kernel density estimate, 
bandwidth chosen with a 10 fold cross validation.
With an offset of $\delta = 10^{-12}$ the symmetric Kullback-Leibler divergence (\mbox{sKL-divergence}) between the distributions of parrots KR and K3 is 0.32 
and between the distributions of parrots FC and GU is 8.17. 
}
\label{fig:KL-div}
\end{figure}
\begin{figure}[H] 
\centering
\includegraphics[width=.59\linewidth]{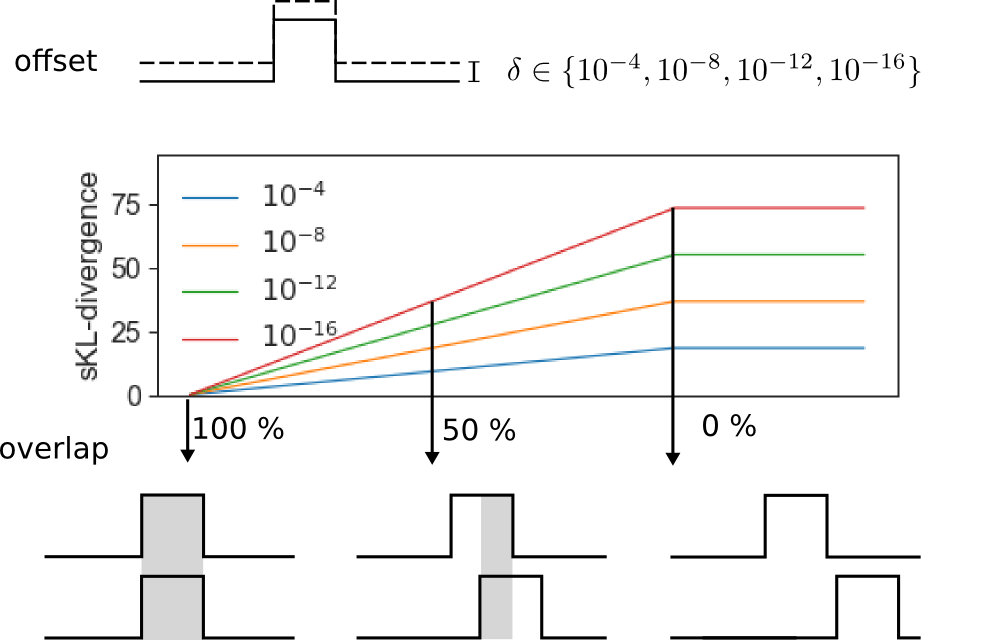}
\caption{
{\bf Effect of the offset $\delta$. }
\mbox{sKL-divergence} between two uniform distributions offset by different values $\delta$ as function of their overlapping region.
In the overlapping region the \mbox{sKL-divergence} decreases linearly with the overlapping fraction. 
The offset $\delta$ sets the scale of the divergence; smaller deltas yielding larger \mbox{sKL-divergences}. 
}
\label{fig:deltas}
\end{figure}

%
The Kullback-Leibler divergence \mbox{(sKL-divergence)} provides a mean for quantifying  how different two probability distributions are. 
%
The more different two probability distributions are, the larger their sKL-divergence; (Fig.~\ref{fig:KL-div}b), 
the more similar the two are, the smaller their sKL-divergence (Fig.~\ref{fig:KL-div}a), being zero only for identical distributions. 
Given two discrete probability distributions $P$ and $Q$ the \mbox{sKL-divergence} is given by
\begin{equation}
D_{\mathrm{sKL}}(P || Q) = \sum_i P_i \, \log \bigg( \frac{P_i}{Q_i} \bigg) + \sum_i Q_i \,  \log \bigg( \frac{Q_i}{P_i} \bigg). 
\label{eq:skl}
\end{equation}
Notice that, when the probability density of either of the distributions is zero while the other is not zero, the \mbox{sKL-divergence} indefines itself. 
In other words we cannot compare probability distributions with zeros. 
For this reason, before comparing distributions we offset them by a small value $\delta$, 
chosen orders of magnitude smaller than the typical values of the distribution. 
Substituting, 
$P(i) \rightarrow P_i + \delta$ and
$Q(i) \rightarrow Q_i + \delta$ in Eq.~\ref{eq:skl} we get, the $\delta$ dependent sKL-divergence, 
\begin{equation}
\begin{split}
D_{\delta \mathrm{sKL}}(P || Q) = 
\sum \limits _i P_i \, \log \bigg( \frac{P_i + \delta}{Q_i + \delta} \bigg) +
\sum \limits _i Q_i \, \log \bigg( \frac{Q_i + \delta }{P_i + \delta} \bigg). 
\end{split}
\end{equation}
%
Because the offset $\delta$ is chosen orders of magnitude smaller than the typical  values of the distributions, its value should not affect overlapping probability distributions.
However, for distributions with non overlapping regions, their \mbox{sKL-divergence} will be affected by the offset $\delta$;  
increasing the \mbox{sKL-divergence} for smaller offset  
while decreasing the \mbox{sKL-divergence} for larger offsets.
\mbox{Figure \ref{fig:deltas}} shows the \mbox{sKL-divergence} between two uniform probability distributions as a function of their overlap, from 100\% to 0\% for for different offsets. 
As expected, the sKL-divergence increases the smaller the overlap is for all values of $\delta$. 
Besides, for a fix value $\delta$, the \mbox{sKL-divergence} decreases linearly with the overlapping fraction. 
So, when comparing distributions is important to keep the same offset, because is the offset that sets the scale of the sKL-divergence.

\section{Results}

We first present the comparisons of the distributions within  the recordings of each dataset 
and then across the recordings of all datasets. 

\subsection{Frogs}

\begin{figure}[h]
\centering
\includegraphics[width=0.5\linewidth]{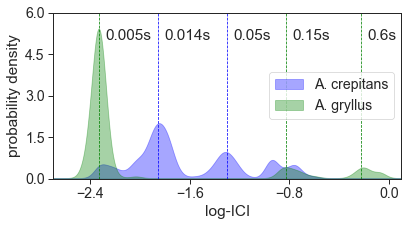}
\caption{
{\bf Distribution of ICI for the two mating songs of frogs {\it Acris gryllus} and {\it Acris crepitans} (Fig.~\ref{fig:frog-call}).} 
Probability distribution function of the logarithm of the inter-pulse intervals (log-ICI) for each frog.
Distribution fitted with a Gaussian KDE, bandwidth chosen with a 10-fold cross validation.
With an an offset $\delta = 10^{-12}$, the symmetric Kullback-Leibler divergence between the two distributions is 15.17.
}
\label{fig:frog-kde}
\end{figure}

The ICIs of the frogs ranged 
from 0.0046 to \mbox{0.25 s} for {\it Acris crepitans}
and from 0.0036 to \mbox{0.87 s} for {\it Acris gryllus}.
The distribution of log-ICI of the frogs has characteristic peaks at different scales (Fig. \ref{fig:frog-kde}). 
The \mbox{sKL-divergence} between both frogs is 20.5 with \mbox{$\delta = 10^{-12}$.}

\subsection{Zebra finches}

\begin{figure}[H]
\centering
\includegraphics[width=.5\linewidth]{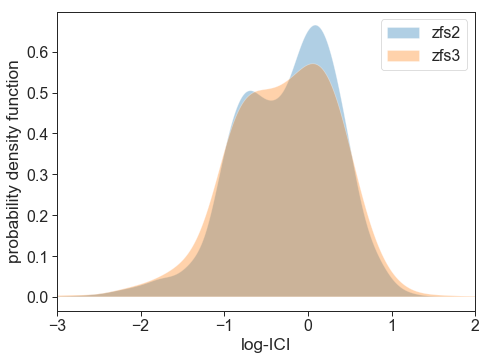}
\caption{
{\bf Comparing the ICIs of a group of four female zebra finches.}
Distributions correspond to two sessions, zfs2 and zfs3, of one hour each with the same four female zebra finches.
With $\delta = 10^{-12}$, the \mbox{sKL-divergence} between the two distributions is 0.04.
}
\label{fig:zf}
\end{figure}

The zebra finches produced overlapping calls which yielded negative ICIs, coming from animals vocalising simultaneously. 
This ICIs were removed previous to analysing and represented 7.6\% of the ICIs for session two and 9.6\% of the ICIs for session three.
The largest ICI was of 22.6 for session two and 3.3 minutes for session three.
The distributions of both days are very similar (Fig.~\ref{fig:zf}), with a \mbox{sKL-divergence} of 0.04 with $\delta = 10^{-12}$.

\subsection{Parrots}

\begin{figure}[h]
\centering
\includegraphics[scale=0.4]{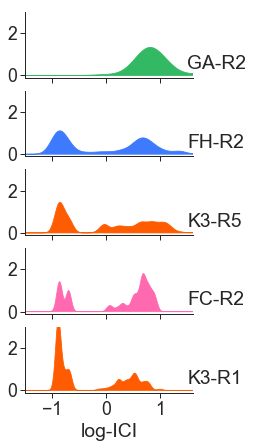}
\put(0,0){\bf a}
\hspace{0.1cm}
\includegraphics[scale=0.48]{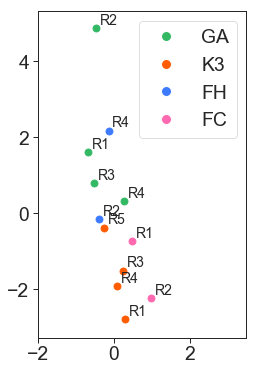}
\put(0,0){\bf b}
\\
\includegraphics[width=0.5\linewidth]{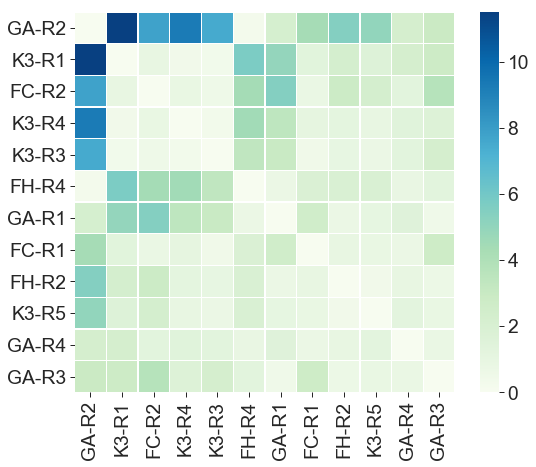}
\put(0,0){\bf c}
\caption{
{\bf Comparing ICIs between parrots of the same species.}
{\bf a}, Probability distribution of the log-ICIs of recordings GA-R2, FH-R2, K3-R5, FC-R2 and K3-R1. 
{\bf b}, Two-dimensional scaling of the symmetric Kullback-Leibler divergence (\mbox{sKL-divergence}) of the distribution of log-ICIs.
{\bf c}, \mbox{sKL-divergence} between the distribution of the log-ICIs of the parrots, with $\delta = 10^{-12}$.
}
\label{fig:compBirds}
\end{figure}

The ICIs of the parrots dataset ranged from 0.04 to 129 s, thus covering five orders of magnitude.
The distributions of most recordings present two probability masses, 
one for small scales at around 0.1 s and 
another one, more broadly distributed, between 1 s and 32 s (Fig.~\ref{fig:compBirds}a).
The probability masses vary from recording to recording (Fig.~\ref{fig:compBirds}a). 
An outlier is recording R2 of bird GA, which is dominated by large ICIs, longer than 1 s (Fig.~\ref{fig:compBirds}a). 
The mean sKL-divergences are $2.5 \pm 2.4$, where the uncertainty is one standard deviation. 
A multidimensional scaling enables to visualise the \mbox{sKL-divergence} in a Euclidean space (Fig.~\ref{fig:compBirds}b-c).

\subsection{Comparison of the timing patterns of all animals}

\begin{figure}[h]
\centering
\includegraphics[width=0.5\linewidth]{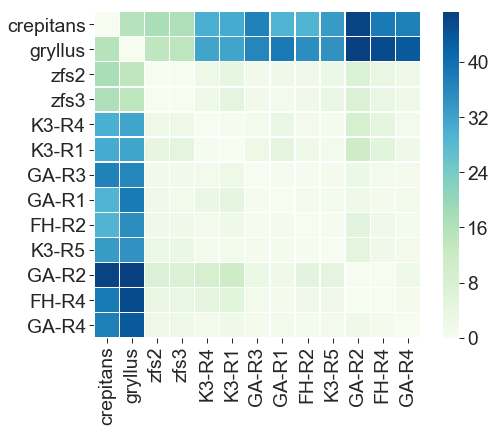}
\put(0,0){\bf a}
\hspace{0.1cm}
\includegraphics[width=0.47\linewidth]{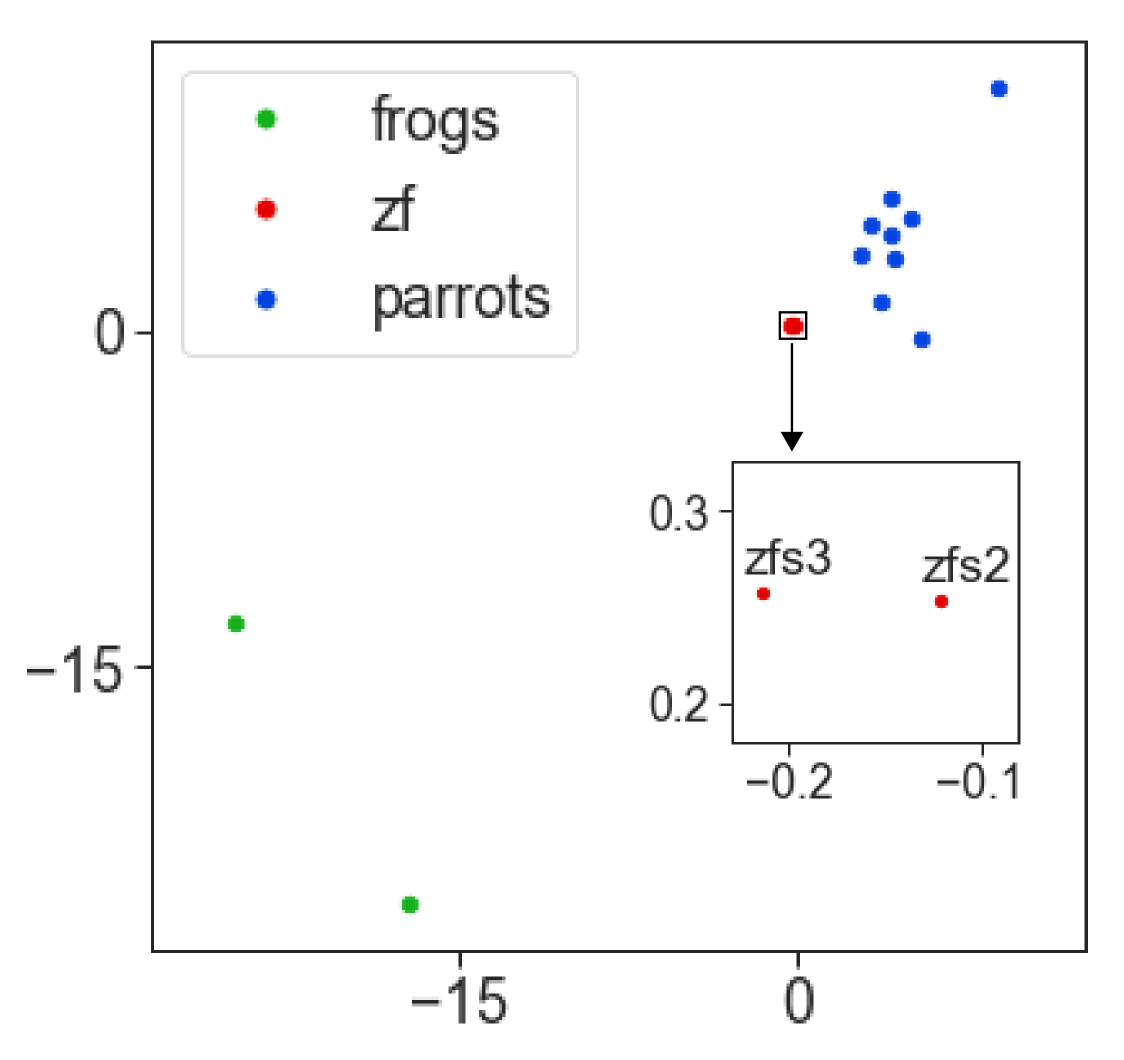} 
\put(0,0){\bf b}

\caption{
{\bf Comparing timing patterns of frogs, zebra finches and parrots. }
{\bf a}, Symmetric Kullback-Leibler divergence (\mbox{sKL-divergence}) between the distributions of inter-call intervals (ICIs) of the tapes of all animals; 
frogs, zebra finches and parrots.
{\bf b}, Two dimensional scaling of the \mbox{sKL-divergence}.
}
\label{fig:comp-all}
\end{figure}

By comparing the distributions of all animals against each other, we obtain 
that closely related animals have more similar distributions than more distantly related animals (Fig. \ref{fig:comp-all}a). 
The average divergences between the recordings of the datasets are  
$16.5 \pm 1.8$ between the frogs and the zebra finches, 
$39.4 \pm 6.3$ between the frogs and the parrots and
$3.4 \pm 1.8$ between the zebra finches and the parrots --
where the standard deviation of the divergences is taken for uncertainty. 
A two dimensional scaling of the distance matrix shows that the zebra finches are so close that is not possible to distinguish their points \mbox{(Fig. \ref{fig:comp-all}b)}. 
Furthermore, the zebra finches are close to the parrots and 
both are quite distant from the frogs, 
which are also distant between each other, but not as distant as to the birds.  
These results agree with the origin of the datasets. 
The most similar points are those from the zebra finches. 
Both points come from the same four animals in the same lab conditions, except for the date and arrangement of the cages.
The points of the parrot cluster do not represent the same individual but come from different individuals of the same species; 
thus these points are expected to be closer than to the rest of the species.
Finally, the frogs {\it Acris gryllus} and {\it Acris crepitans} are neither the same individual nor of the same species but share the genus; 
thus differences in their timing patterns are expected, however not as big as the differences form entirely different animals.

\subsection{Comparison with Fourier analysis}

Fourier analysis captures the periodic structure of a time series by decomposing it into sums of sinusoid  functions. 
To compare our method with Fourier analysis, 
instead of using the distribution of log-ICIs as timing features, 
we use the power spectral density (PSD) as timing features 
and compare spectral densities with the sKL-divergence. 
Because we are interested in the temporal structure of the vocalisations, 
we define a binary time series from the onset times of the calls, 
with ones at the onset times and zeros elsewhere.  
This way, 
the high frequencies of the spectrum will be associated with 
periodicities in the short inter-onset intervals (IOI)
and 
the low frequencies of the spectrum will be associated with 
periodicities in the large inter-onset intervals and other large periodicities of the signal. 
Because Fourier analysis acts over a time series, 
it cannot handle ICIs.
Our method, on the other hand, is more flexible since it can 
either operate over the ICIs or the IOIs. 
%

To make a fair comparison between our method and the Fourier analysis we should consider the same rage of temporal structures. 
If with our method we resolve IOIs as small as a cent of a second, Fourier analysis should also have the capability to resolve such small scales and the same for large scales. 
The sampling rate of the time series determines the frequency resolution of the spectrum and therefore the smallest inter-onset interval that we can measure. 
Let $IOI_{\mbox{min}}$ be the smallest IOI we can resolve, 
then, because of the Nyquist frequency we chose the sampling rate to be $2/IOI_{\mbox{min}}$. 
%
On the other side of the spectrum, 
large periodicities are limited by the size of the Fourier transform window. 
We want to be able to detect large IOI, so we select the largest possible window which is the length of the time series. 
However, because in general the recordings have different durations this yields spectrums of different frequency ranges and different resolutions, 
thereby difficulting their comparison. 
To be able to resolve for large periods and have spectrums within the same range and resolution, 
we zero pad the time series and use a large window -- at least as long as the longest time series being compared. 

\begin{figure*}[!htbp]

\centering
\includegraphics[width=0.99\textwidth]{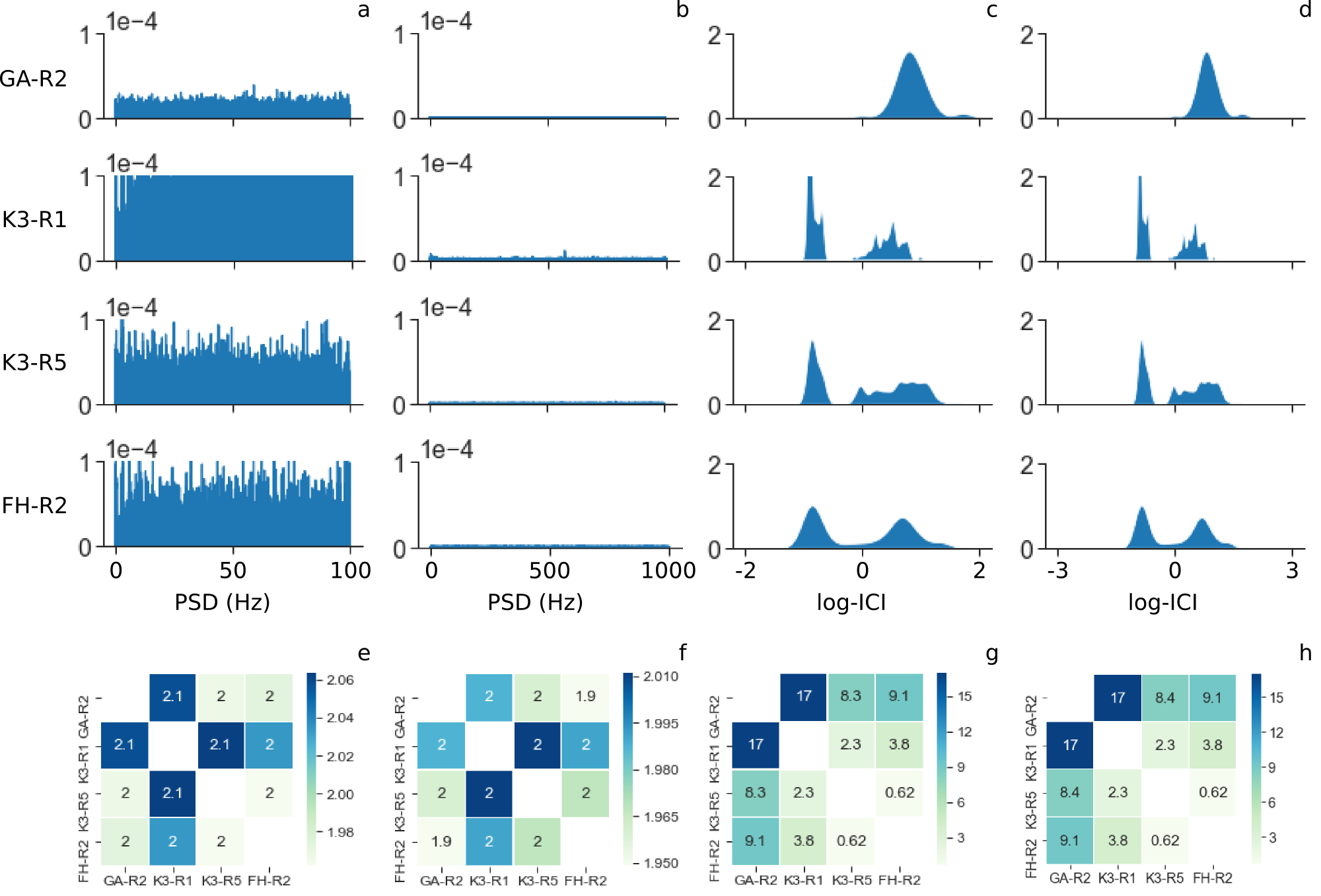}

\caption{
{\bf Comparison of our method with Fourier analysis. }
The top rows shows the features of four parrot tapes: GA-R2, K3-R1, K3-R5 are FH-R2 and the bottom row shows the sKL-divergence matrix from the above features. 
{\bf a}, Power spectral density the onset times of each recording in the range 0 to 100 Hz. 
Time series defined with a sampling rate of 200 Hz,  
and spectrum computed zero padding the signal with a window of $2^{18}$ points. 
{\bf b}, Power spectral density the onset times of each recording in the range 0 to  1000 Hz.  
Time series defined with a sampling rate of 2000 Hz,  
and spectrum computed zero padding the signal with a window of $2^{21}$ points. 
{\bf c}, KDE in the range 0.01 s to 100 s evaluated on $2^{18}$ points. 
{\bf b}, KDE in the range 0.001 s to 1000 s evaluated on $2^{21}$ points. 
}
\label{fig:KDE_vs_FFT}
\end{figure*}
%
%
%
Comparing both methods on a subset of four parrot recordings (Fig.~\ref{fig:KDE_vs_FFT}) we find that the power spectral density in not capturing the information as well as our method with the KDE of the log-ICI. 
The first thing we observe is that the range of variation of the sKL-divergences  computed with the PSD is much smaller than the one computed with our method. 
The span of the sKL-divergences computed with the PSD is 0.2 and 
the span of the sKL-divergences computed with the PSD is 11. 
A second and more crucial issue comes when changing parameters used for computing the power spectral density (Fig.~\ref{fig:KDE_vs_FFT}a-b). 
We tried two parameter sets, 
one set with a sampling rate of 200 Hz and Fourier window of $2^{18}$ which yields a spectrum in the range 0 to 100 Hz 
and 
another set with a sampling rate of 2000 Hz and Fourier window of $2^{21}$ which yields a spectrum in the range 0 to 1000 Hz.  
%
These sets of parameters yields very different feature representations that result in two sKL-divergences that do not preserve the relative distances.  
With the first set of parameters 
tape FH-R2 is more similar to tape K3-R5 than to tape GA-R2 
whereas with the second set of parameters 
tape FH-R2 is more similar to tape GA-R2 than to tape K3-R5. 
Both parameter sets are valid; 
the second can resolve smaller and larger periods, 
but this periods are beyond the range of variation of the IOI. 

\section{Discussion}

We proposed a novel method for investigating timing in animal vocalisations. 
The method consist of parametrising timing as the distribution of ICIs 
and comparing the distributions with the sKL-divergence. 
We used this method to compare distributions of ICIs in three datasets:  
(1) two closely related frog species, 
(2) a group of four zebra finches recorded on two different days 
and (3) 12 recordings from four parrots of the same species. 
We first compared the distributions 
within the recordings of each dataset 
and then across the recordings of all datasets. 
The within dataset comparisons show that 
both frog species have very distant distributions of ICIs, 
with a sKL-divergence of 15.17; 
the zebra finches have very similar distributions, 
with a sKL-divergence of 0.4 
and the parrots have sKL-divergence ranging from 0.3 to 11.5  
with an average of $2.4$ with a standard deviation of $2.2$. 
All together, the distributions of the parrots, 
on average differed more than those of the zebra finches 
and less than those of the frogs. 
When comparing the divergences across datasets we obtain  
average divergences that are larger than the divergences within dataset;  
$16.5 \pm 1.8$ between the frogs and the zebra finches, 
$39.4 \pm 6.3$ between the frogs and the parrots and
$3.4 \pm 1.8$ between the zebra finches and the parrots. 
So, the ICIs of the birds --zebra finches and parrots-- 
are more similar to each other than they are to the frogs. 
This follows from the fact that 
the typical ICIs of the frogs 
are orders of magnitude smaller 
than the typical ICIs of the birds. 
Our method can process and compare ICIs of different time scales -- not only being suitable for investigating ICIs of any taxa, 
but also for comparing ICIs of different taxa against each other as we did here.

Different time scales not only arise when comparing different taxa but within a taxon, ICIs are distributed multimodally over several time scales.
Because of this wide range and
 because the distribution of ICIs is not centred \cite{frasier2017automated}, 
 statistics like the mean, the median and the mode are of little help for summarising the ICIs. 
A common way of handling different time scales is by ad hoc defining categories of ICIs. 
For example, groups of calls \cite{nevo1985evolutionary},
from which intra-group intervals 
and inter-group intervals are estimated separately \cite{micancin2014allometric}. 
However, observer-based steps like this, not only are time consuming but prone to observer bias. 
Here, we turned around the different scales problem by log transforming the ICIs. 
The log transform enabled to investigate timing patterns at different time scales simultaneously. 
Moreover, the different ICI scales emerge in our method through the modes in the distribution of log-ICIs. 

Frogs, zebra finches and parrots are very different species so it is not surprising that their timing patterns also differ widely.
Additionally, the distribution of ICIs has different meanings for each dataset.
The frogs produce pulses and were recorded in isolation, 
so the ICIs are in fact inter-pulse intervals of one frog.
The zebra finches were recorded in group, 
so the ICIs may come from calls emitted by the same or different animals. 
The parrot recordings are from nesting vocalisations, 
a context in which only one male bird is vocalising, 
so the ICIs come from call emitted by a single parrot.
The aim of comparing timing patterns of very different species with conceptually different ICIs, 
is to illustrate the reasoning behind the method.
%
%
This approach may be more valuable for comparable samples like systematic recordings of closely related species.
%
  
There are temporal emissions patterns that our method would not be able to differentiate. 
Consider for instance two recordings with the same distribution of ICIs but different ICI order. 
One may implement this by randomising the ICIs.
Because the distribution of ICIs is the same,  
our method would not be able to differentiate them.  
While this is a limitation to keep in mind, 
our method was proven useful for distinguishing timing in vocalisations of different animals.
The distribution of ICIs is capturing the fingerprint of the temporal emission patterns well enough for distinguishing different species as observed here.

\subsection{Technical details of our method}


The method here proposed depends on three parameters: 
the bandwidth and the kernel of the KDE and the offset $\delta$. 
%
%
Below we discuss the relevance of these parameters to the method and the rational for choosing them. 

Both, bandwidth and kernel, were selected with a ten fold cross validation.
Cross validation enables to brute force evaluate a set of parameters and choose the ones that maximises a scoring criterion. 
Evaluation is done by splitting a dataset into ten train-test datasets and scoring the KDE fitted with the train set as likelihood of the test set. 
Because the distribution often has several modes, choosing the bandwidth using cross validation yields better scores than choosing with 
Silverman's rule of thumb \cite{silverman2018density}.

To avoid zeros in the sKL-divergence, we offsetted the probability distributions by a $\delta = 10^{-12}$. 
%
%
When comparing overlapping distributions, the value of $\delta$ plays no role because the values of the distribution will be much larger than $\delta$. 
However, in regions with no overlap, 
the offset will define the scale of the distance.
So, when comparing divergences, one should use the same offset. 
Consequently, when reporting sKL-divergences it is important to do so together with the value of $\delta$. 

The simplicity of the presented method enables to apply it to other variables. 
The backbone of the method consist of 
fitting a distribution with a kernel density estimate and 
comparing distributions with the symmetric KL-divergence. 
This backbone may be used to compare other continuous variable parameters, such as the distribution of call lengths.

\subsection{Comparison with other methods}

We also used Fourier's PSD as timing feature and compared different PSD with the sKL-divergence. 
However, we obtained that this approach cannot handle recordings with different sample points robustly. 
Fourier analysis is highly dependent on the window size and the sampling rate used to binarise the signal. 
Both parameters are crucial since they determine the resolution of the PSD, yet there is no a priory criterion for setting these parameters. 
Moreover, choosing different parameter, yields inconsistent results. 
When widening the analysis region, distributions that previously were similar can become distant. 
A robust method should preserve the relative distances to this kind of parameters. 
While this is true for our method, is not for the Fourier features (Fig.~\ref{fig:KDE_vs_FFT}). 
This limits the applicability of the PSD to particular time scale. 
Our method can compare distributions of ICIs of different time scales as was successfully illustrated with the frogs and the parrots, whose distributions span over more than four orders of magnitude.

Besides Fourier analysis other methods have been proposed to investigate temporal structures in animal vocalisations \cite{ravignani2017measuring}. 
Phase plots \cite{rothenberg2014investigation} for instance 
yield visualisations that highlight the temporal structures as geometric patterns and so require trained humans to tell similar and different patterns apart. 
The method here proposed does not depend on human decisions, but measures similarity objectively and in a reproducible way. 

\subsection{Conclusion}

We presented a simple, flexible and robust method to quantitatively investigate timing in animal vocalisations based in the distribution of log-ICIs.  
Unlike Fourier analysis, our method is robust to changes in the analysis range. 
Additionally, we provided guidelines for choosing the parameters of the method (bandwidth of the KDE, $\delta$). 
%
%
The proposed method is not limited to one species but can be used to investigate the timing patterns of any taxa as illustrated here comparing the ICIs of frogs, zebra finches and parrots. 
The simplicity of the method allows the same reasoning to be used to investigate other continuous variables such as call duration or even other continuous variables beyond the scope of bioacoustics. 
Investigate timing in an objective an reproducible manner 
can contribute to taxonomy, support conservation efforts by helping monitoring animals in the wild and may shed light onto the origins of timing structures in animal vocal communication.

\subsection{Acknowledgements}

We thank Andrea Raviganani, Dan Stowell and Aline Viol for discussions.
We also thank Katherine Renton for supporting ACM-M to obtain the vocal recordings of Lilac-crowned Parrots.
This work was 
supported by the Deutsche Forschungsgemeinschaft (DFG, German Research Foundation) under Germany´s Excellence Strategy – EXC-2068 – 390729961 – Cluster of Excellence Physics of Life and the Cluster of Excellence Center for Advancing Electronics at TU Dresden, 
CONACYT Mexico and the Max Planck Society.

\section{References}

\bibliographystyle{ieeetr} 

\bibliography{/home/florencia/profesjonell/bibliography/bibliography.bib}

\begin{thebibliography}{10}

\bibitem{mahurin2008chick}
E.~J. Mahurin and T.~M. Freeberg, ``Chick-a-dee call variation in carolina
  chickadees and recruiting flockmates to food,'' {\em Behavioral Ecology},
  vol.~20, no.~1, pp.~111--116, 2008.

\bibitem{slocombe2006food}
K.~E. Slocombe and K.~Zuberb{\"u}hler, ``Food-associated calls in chimpanzees:
  responses to food types or food preferences?,'' {\em Animal Behaviour},
  vol.~72, no.~5, pp.~989--999, 2006.

\bibitem{freeberg2012geographic}
T.~M. Freeberg, ``Geographic variation in note composition and use of
  chick-a-dee calls of carolina chickadees (poecile carolinensis),'' {\em
  Ethology}, vol.~118, no.~6, pp.~555--565, 2012.

\bibitem{sayigh1999individual}
L.~S. Sayigh, P.~L. Tyack, R.~S. Wells, A.~R. Solow, M.~D. Scott, and
  A.~Irvine, ``Individual recognition in wild bottlenose dolphins: a field test
  using playback experiments,'' {\em Animal behaviour}, vol.~57, no.~1,
  pp.~41--50, 1999.

\bibitem{charrier2009vocal}
I.~Charrier, B.~J. Pitcher, and R.~G. Harcourt, ``Vocal recognition of mothers
  by australian sea lion pups: individual signature and environmental
  constraints,'' {\em Animal Behaviour}, vol.~78, no.~5, pp.~1127--1134, 2009.

\bibitem{ford1989acoustic}
J.~K.~B. Ford, ``Acoustic behaviour of resident killer whales (orcinus orca)
  off vancouver island, british columbia.,'' {\em Canadian Journal of Zoology},
  vol.~67, pp.~727--745, 1989.

\bibitem{attard2010vocal}
M.~R. Attard, B.~J. Pitcher, I.~Charrier, H.~Ahonen, and R.~G. Harcourt,
  ``Vocal discrimination in mate guarding male australian sea lions:
  familiarity breeds contempt,'' {\em Ethology}, vol.~116, no.~8, pp.~704--712,
  2010.

\bibitem{arcadi1996phrase}
A.~C. Arcadi, ``Phrase structure of wild chimpanzee pant hoots: patterns of
  production and interpopulation variability,'' {\em American Journal of
  Primatology}, vol.~39, no.~3, pp.~159--178, 1996.

\bibitem{seyfarth1980vervet}
R.~M. Seyfarth, D.~L. Cheney, and P.~Marler, ``Vervet monkey alarm calls:
  semantic communication in a free-ranging primate,'' {\em Animal Behaviour},
  vol.~28, no.~4, pp.~1070--1094, 1980.

\bibitem{casar2012evidence}
C.~C{\"a}sar, R.~W. Byrne, W.~Hoppitt, R.~J. Young, and K.~Zuberb{\"u}hler,
  ``Evidence for semantic communication in titi monkey alarm calls,'' {\em
  Animal Behaviour}, vol.~84, no.~2, pp.~405--411, 2012.

\bibitem{nevo1985evolutionary}
E.~Nevo and R.~R. Capranica, ``Evolutionary origin of ethological reproductive
  isolation in cricket frogs, acris,'' in {\em Evolutionary biology},
  pp.~147--214, Springer, 1985.

\bibitem{bohn2008syllable}
K.~M. Bohn, B.~Schmidt-French, S.~T. Ma, and G.~D. Pollak, ``Syllable
  acoustics, temporal patterns, and call composition vary with behavioral
  context in mexican free-tailed bats,'' {\em The Journal of the Acoustical
  Society of America}, vol.~124, no.~3, pp.~1838--1848, 2008.

\bibitem{micancin2014allometric}
J.~P. Micancin and R.~H. Wiley, ``Allometric convergence, acoustic character
  displacement, and species recognition in the syntopic cricket frogs acris
  crepitans and a. gryllus,'' {\em Evolutionary Biology}, vol.~41, no.~3,
  pp.~425--438, 2014.

\bibitem{simard2010depth}
P.~Simard, A.~L. Hibbard, K.~A. McCallister, A.~S. Frankel, D.~G. Zeddies,
  G.~M. Sisson, S.~Gowans, E.~A. Forys, and D.~A. Mann, ``Depth dependent
  variation of the echolocation pulse rate of bottlenose dolphins (tursiops
  truncatus),'' {\em The Journal of the Acoustical Society of America},
  vol.~127, no.~1, pp.~568--578, 2010.

\bibitem{evans1994food}
C.~S. Evans and P.~Marler, ``Food calling and audience effects in male
  chickens, gallus gallus: their relationships to food availability, courtship
  and social facilitation,'' {\em Animal Behaviour}, vol.~47, no.~5,
  pp.~1159--1170, 1994.

\bibitem{wells1984vocal}
K.~D. Wells and J.~J. Schwartz, ``Vocal communication in a neotropical
  treefrog, hyla ebraccata: advertisement calls,'' {\em Animal Behaviour},
  vol.~32, no.~2, pp.~405--420, 1984.

\bibitem{matos2005performing}
R.~J. Matos and I.~Schlupp, ``Performing in front of an audience: signalers and
  the social environment,'' {\em Animal communication networks}, pp.~63--83,
  2005.

\bibitem{payne2003elephant}
K.~B. Payne, M.~Thompson, and L.~Kramer, ``Elephant calling patterns as
  indicators of group size and composition: the basis for an acoustic
  monitoring system,'' {\em African Journal of Ecology}, vol.~41, no.~1,
  pp.~99--107, 2003.

\bibitem{sun2005anthropogenic}
J.~W. Sun and P.~M. Narins, ``Anthropogenic sounds differentially affect
  amphibian call rate,'' {\em Biological conservation}, vol.~121, no.~3,
  pp.~419--427, 2005.

\bibitem{kaiser2009effect}
K.~Kaiser and J.~L. Hammers, ``The effect of anthropogenic noise on male
  advertisement call rate in the neotropical treefrog, dendropsophus
  triangulum,'' {\em Behaviour}, vol.~146, no.~8, pp.~1053--1069, 2009.

\bibitem{potvin2011geographically}
D.~A. Potvin, K.~M. Parris, and R.~A. Mulder, ``Geographically pervasive
  effects of urban noise on frequency and syllable rate of songs and calls in
  silvereyes (zosterops lateralis),'' {\em Proceedings of the Royal Society of
  London B: Biological Sciences}, vol.~278, no.~1717, pp.~2464--2469, 2011.

\bibitem{cranford1999sperm}
T.~W. Cranford, ``The sperm whale's nose: Sexual selection on a grand scale?
  1,'' {\em Marine mammal science}, vol.~15, no.~4, pp.~1133--1157, 1999.

\bibitem{gero2016individual}
S.~Gero, H.~Whitehead, and L.~Rendell, ``Individual, unit and vocal clan level
  identity cues in sperm whale codas,'' {\em Royal Society Open Science},
  vol.~3, no.~1, p.~150372, 2016.

\bibitem{stark1998quantitative}
R.~D. Stark, D.~J. Dodenhoff, and E.~V. Johnson, ``A quantitative analysis of
  woodpecker drumming,'' {\em Condor}, pp.~350--356, 1998.

\bibitem{mann2009comparative}
N.~I. Mann, K.~A. Dingess, K.~F. Barker, J.~A. Graves, and P.~J. Slater, ``A
  comparative study of song form and duetting in neotropical thryothorus
  wrens,'' {\em Behaviour}, vol.~146, no.~1, pp.~1--43, 2009.

\bibitem{weilgart1997group}
L.~Weilgart and H.~Whitehead, ``Group-specific dialects and geographical
  variation in coda repertoire in south pacific sperm whales,'' {\em Behavioral
  Ecology and Sociobiology}, vol.~40, no.~5, pp.~277--285, 1997.

\bibitem{kershenbaum2014acoustic}
A.~Kershenbaum, D.~T. Blumstein, M.~A. Roch, {\c{C}}.~Ak{\c{c}}ay, G.~Backus,
  M.~A. Bee, K.~Bohn, Y.~Cao, G.~Carter, C.~C{\"a}sar, {\em et~al.}, ``Acoustic
  sequences in non-human animals: a tutorial review and prospectus,'' {\em
  Biological Reviews}, vol.~91, no.~1, pp.~13--52, 2014.

\bibitem{stowell_2016}
D.~Stowell, ``Zebra finch group calling zf4f,'' Jan 2016.

\bibitem{renton1999nesting}
K.~Renton and A.~Salinas-Melgoza, ``Nesting behavior of the lilac-crowned
  parrot,'' {\em The Wilson Bulletin}, pp.~488--493, 1999.

\bibitem{montes2016contextual}
A.~C. Montes-Medina, A.~Salinas-Melgoza, and K.~Renton, ``Contextual
  flexibility in the vocal repertoire of an amazon parrot,'' {\em Frontiers in
  zoology}, vol.~13, no.~1, p.~40, 2016.

\bibitem{gramacki2018nonparametric}
A.~Gramacki, {\em Nonparametric Kernel Density Estimation and Its Computational
  Aspects}.
\newblock Springer, 2018.

\bibitem{pedregosa2011scikit}
F.~Pedregosa, G.~Varoquaux, A.~Gramfort, V.~Michel, B.~Thirion, O.~Grisel,
  M.~Blondel, P.~Prettenhofer, R.~Weiss, V.~Dubourg, {\em et~al.},
  ``Scikit-learn: Machine learning in python,'' {\em The Journal of Machine
  Learning Research}, vol.~12, pp.~2825--2830, 2011.

\bibitem{frasier2017automated}
K.~E. Frasier, M.~A. Roch, M.~S. Soldevilla, S.~M. Wiggins, L.~P. Garrison, and
  J.~A. Hildebrand, ``Automated classification of dolphin echolocation click
  types from the gulf of mexico,'' {\em PLoS computational biology}, vol.~13,
  no.~12, p.~e1005823, 2017.

\bibitem{silverman2018density}
B.~W. Silverman, {\em Density estimation for statistics and data analysis}.
\newblock Routledge, 2018.

\bibitem{ravignani2017measuring}
A.~Ravignani and P.~Norton, ``Measuring rhythmic complexity: A primer to
  quantify and compare temporal structure in speech, movement, and animal
  vocalizations,'' {\em Journal of Language Evolution}, p.~lzx002, 2017.

\bibitem{rothenberg2014investigation}
D.~Rothenberg, T.~C. Roeske, H.~U. Voss, M.~Naguib, and O.~Tchernichovski,
  ``Investigation of musicality in birdsong,'' {\em Hearing research},
  vol.~308, pp.~71--83, 2014.

\end{thebibliography}

\end{document}